\newcommand{\gtap}{\mathrel{\hbox{\rlap{\lower.55ex \hbox {$\sim$}}
                   \kern-.3em \raise.4ex \hbox{$>$}}}}
\newcommand{\ltap}{\mathrel{\hbox{\rlap{\lower.55ex \hbox {$\sim$}}
                   \kern-.3em \raise.4ex \hbox{$<$}}}}

\documentclass{iaus}
\usepackage{psfig,graphicx}

\title[X-ray sources in globular clusters] 
{Observational evidence for the origin of X-ray sources in globular clusters}

\author[Frank Verbunt, Dave Pooley, Cees Bassa]   
{Frank Verbunt$^1$,\,%
Dave Pooley$^2$ \break \and Cees Bassa$^3$}

\affiliation{$^1$Astronomical Institute, Postbox 80.000,
3508 TA Utrecht, the Netherlands
 \break email: verbunt@astro.uu.nl\\[\affilskip]
$^2$Dept. of Astronomy, University of Wisconsin-Madison
Madison WI 53706-1582, U.S.A.
\break email: dave@astro.wisc.edu\\[\affilskip]
$^3$Physics Department, McGill University, Montreal, QC H3A 2T8 Canada
\break email: bassa@physics.mcgill.ca}

\pubyear{2007}
\volume{246}  
\pagerange{1--10}
\date{\today}
\jname{Dynamical evolution of dense stellar systems}
\editors{E. Vesperini}
\begin{document}

\maketitle

\begin{abstract}
Low-mass X-ray binaries, recycled pulsars, cataclysmic variables and
magnetically active binaries are observed as X-ray sources in globular 
clusters. We discuss the classification of these systems, and find that
some presumed active binaries are brighter than expected. We discuss a new
statistical method to determine from observations how the formation 
of X-ray sources depends on the number of
stellar encounters and/or on the cluster mass.
We show that cluster mass is not a proxy for the encounter number,
and that optical identifications are essential in proving the presence of
primordial binaries among the low-luminosity X-ray sources.
\keywords{X-ray sources, globular clusters, stellar encounters}
\end{abstract}

\firstsection 
\section{Introduction}

The first celestial maps in X-rays, in the early 1970s, show that
globular clusters harbour more X-ray sources than one would expect
from their mass. As a solution to this puzzle it was suggested that
these bright ($L_x\gtap10^{36}$\,erg/s) X-ray sources, binaries in
which a neutron star captures mass from a companion star, are formed
in close stellar encounters. A neutron star can be caught by a
companion in a tidal capture, or it can take the place of a star in a
pre-existing binary in an exchange encounter. Verbunt \&\ Hut (1987)
showed that the probability of a cluster to harbour a bright
X-ray source indeed scales with the number of stellar encounters
occurring in it; whereas a scaling with mass does not explain the
observations.

With the \textit{Einstein} satellite a dozen less luminous
 ($L_x\ltap10^{35}$\,erg/s) X-ray sources were discovered in the early
1980. \textit{ROSAT} enlarged this number to some 55, and now thanks
to \textit{Chandra} we know hundreds of dim X-ray sources in
globular clusters. The nature and origin of these dim sources is 
varied. Those containing neutron stars, i.e.\ the quiescent low-mass 
X-ray binaries in which a neutron star accretes mass from its companion
at a low rate and the recycled or millisecond radio pulsars,
have all formed in processes involving close stellar encounters.
The magnetically active binaries, on the other hand, are most
likely primordial binaries, with stars that are kept in rapid
rotation via tidal interaction. Cataclysmic variables are binaries
in which a white dwarf accretes matter from a companion. In globular
clusters they may arise either via stellar encounters, or from
primordial binaries through ordinary binary evolution -- this
is expected to depend on the mass and density of the globular cluster.

In this paper we  describe the classification and identification
of the dim sources in Section\,2, and make some remarks on the theory
of their formation in Section\,3. In Section\,4 we will discuss a
new, and in our view more accurate, way to compare the numbers of these
sources with theoretical predictions.

\begin{figure}
\centerline{
\parbox[b]{0.55\columnwidth}{\psfig{figure=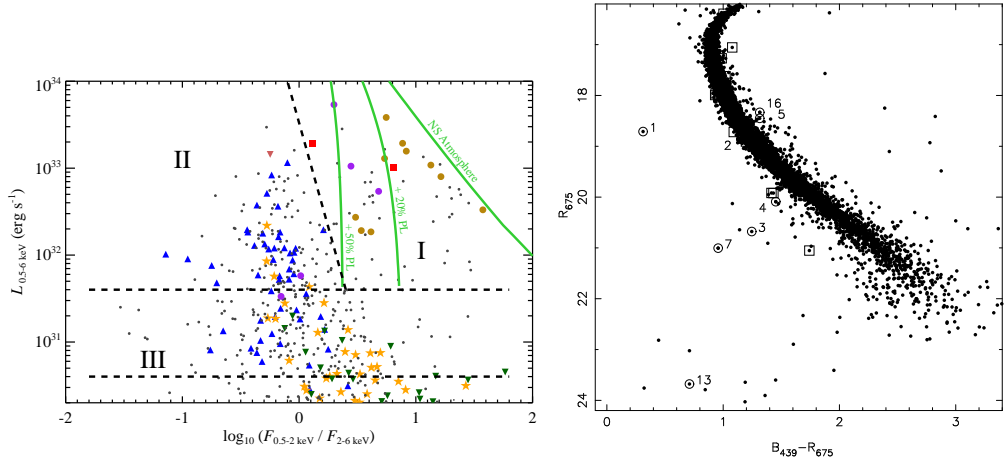,width=0.54\columnwidth,clip=t}}
\parbox[b]{0.45\columnwidth}{\psfig{figure=verbuntf1b.ps,width=0.44\columnwidth,clip=t}}
}

\caption{Left: X-ray hardness-luminosity diagram for dim sources
in globular clusters. I: quiescent low-mass X-ray binaries,
II: cataclysmic variables III: cataclysmic variables and magnetically
active binaries. From Pooley \&\ Hut (2006). 
Right: Colour-magnitude diagram of NGC\,6752 on the basis of
HST-WFPC2 data; objects within X-ray position error circles are
marked. Left of the main sequence we find cataclysmic variables,
above it active binaries. Updated from Pooley et al.\ (2002a).\label{xcol}}
\end{figure}

\section{Clasification and identification \label{idclas}}

Work on the dim sources is progressing along various lines.
Grindlay and coworkers study one cluster, 47\,Tuc,
in great detail (Grindlay et al.\ 2001, Edmonds et al.\ 2003,
Heinke et al.\ 2005). 
Webb and coworkers use XMM to obtain high-quality
X-ray spectra (e.g.\ Webb et al.\ 2006, Servillat this meeting). 
Dim sources are also found in clusters in which
individual sources are the main target, such as Terzan\,1 and 5,
and M\,28 (Wijnands et al.\ 2002, Heinke et al.\ 2003, 
Becker et al. 2003).
Lewin initiated a large program to observe clusters with
very different central densities and core radii, and thereby to provide
material for tests on the dependence on these properties
of the numbers of dim sources.
Further references to all this work may be found in the
review by Verbunt \&\ Lewin (2006); and in the remainder of this 
Section.

The first classification of the dim sources may be made on the basis
of the {\bf X-ray properties only} (Fig.\,\ref{xcol}). The brightest
sources in the 0.5-2.5\,keV band, at $L_x\gtap10^{32}$ erg/s, tend to
be quiescent low-mass X-ray binaries. To better use the Chandra range,
one may also select the brightest sources in the 0.5-6.0\,keV band,
and select soft sources, with a high ratio of fluxes below and above
e.g. 2\,keV: $f_\mathrm{0.5-2.0keV}/f_\mathrm{2.0-6.0keV}\gtap1$.  Such
sources also are mostly quiescent low-mass X-ray binaries.  Between
$10^{31}$ and $10^{32}$ erg/s most sources are cataclysmic variables,
especially when they have hard spectra
$f_\mathrm{0.5-2.0keV}/f_\mathrm{2.0-6.0keV}<1$. The faintest sources
include magnetically active binaries, often with soft X-ray spectra.
For many faint sources the number of counts is too low to decide on
the hardness of the spectrum.  

The second step in classification can be made when {\bf
identification} with a source {\bf at other wavelengths} is
made. Positional coincidence of an X-ray source with the accurate
radio position of a millisecond pulsar provides a reliable
identification and classification.  Positional coincidence with
optical sources is only significant if the highest possible
astrometric accuracy is used to limit the number of possible
counterparts (e.g.\ Bassa et al.\ 2004).  The position of these
possible counterparts in a colour-magnitude diagram is then used to
select the probable counterparts. Cataclysmic variables are bluer than
the main sequence stars, and magnetically active binaries may lie
above the main sequence. Systems on the main sequence cannot be
unambiguously classified: they may either be cataclysmic variables in
which the optical flux is dominated by the donor star, or
main-sequence binaries with unequal masses whose optical light is
dominated by the brighter star.  If a periodicity is found in the
X-rays that corresponds to a period at another wavelength, e.g. the
pulse period of a pulsar or the orbital period of a binary,
identification and clasification are secured simultaneously
(e.g.\ Ferraro et al.\ 2001).

\begin{figure}
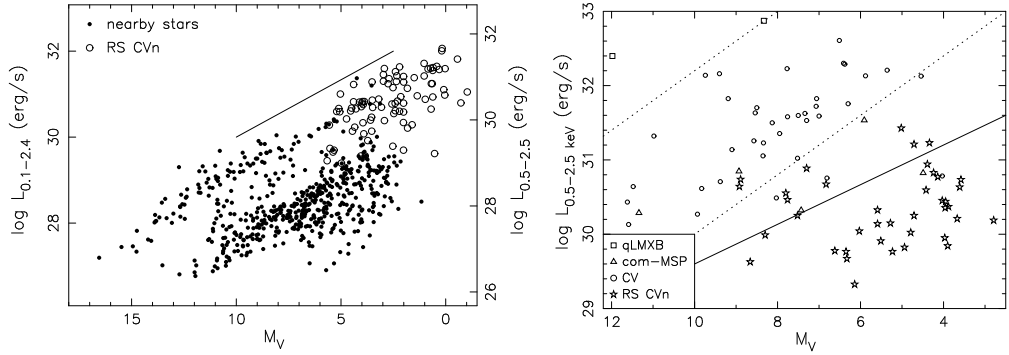

\parbox[b]{0.54\columnwidth}{\psfig{figure=verbuntf2a.ps,width=0.53\columnwidth,clip=t}}
\parbox[b]{0.46\columnwidth}{\psfig{figure=verbuntf2b.ps,width=0.45\columnwidth,clip=t}}

\caption{Left: X-ray luminosity as a function of absolute visual
magnitude for nearby stars ({\em selected} from H\"unsch et al.\ 1999,
for details see Verbunt 2001) and for RS CVn systems (from Dempsey et
al.\ 1993). The upper bound Eq.\,\ref{lxmax} is indicated with a solid
line. We convert the X-ray fluxes in the 0.1-2.4 keV range (scale on
the left) to the 0.5-2.5 keV range by multiplication with 0.4 (scale
on the right).  Right: X-ray luminosity as a function of absolute
visual magnitude, for dim X-ray sources in globular clusters. The
assumed separatrices Eqs.  \ref{sepa},\ref{sepb} are indicated with
dotted lines, the upper bound Eq.\,\ref{lxmax}
with a solid line. It is seen that some X-ray sources classified as
active binaries in globular clusters are well above this
bound.\label{lxmv}}
\end{figure}

A very useful discriminant in X-ray astronomy in general is the
X-ray to optical flux ratio. In the case of globular clusters we
can use the known distance to determine the {\bf optical to X-ray
luminosity ratio} (Fig\,\ref{lxmv}).
On the basis of in particular the extensive
data on 47\,Tuc (Edmonds et al.\ 2003 and references therein) one 
finds that the lines of constant optical to X-ray luminosity ratio
\begin{equation}
\log L_\mathrm{0.5-2.5keV}\mathrm{(\,erg/s)}=36.2 -0.4M_\mathrm{V}
\label{sepa}\end{equation}
separates the quiescent low-mass X-ray binaries above it from the
cataclysmic variables below. The line
\begin{equation}
\log L_\mathrm{0.5-2.5keV}\mathrm{(\,erg/s)}=34.0 -0.4M_\mathrm{V}
\label{sepb}\end{equation}
roughly separates the cataclysmic variables from the magnetically
active binaries.

This latter separatrix leads to a surprise when one compares it with
the X-ray luminosities of nearby stars and of known magnetically
active binaries, i.e.\ RS CVn systems, near the Sun.  For
main-sequence stars in the solar neighbourhood, the X-ray luminosity
increases with the rotation speed, up to an upper bound given
approximately by
\begin{equation}
\log L_\mathrm{0.5-2.5keV}\mathrm{(\,erg/s)}=32.3 -0.27M_\mathrm{V}
\label{lxmax}\end{equation}
as illustrated in Fig.\,\ref{lxmv} (left). 
This bound is lower than the separatrix given by Eq.\,\ref{sepb},
especially for brighter stars. This would imply that active
binaries in globular clusters can have higher X-ray luminosities
than similar binaries near the Sun. We suggest, however, that
the classification must be reinvestigated, and that some of
these objects are cataclysmic variables. The absence of
the blue colour expected for a cataclysmic variable (see Fig.\,\ref{xcol})
then requires explanation --  e.g.\ as a consequence of the 
non-simultaneous measurements at different colours combined with
source variability.\footnote{In his contribution to this meeting,
Christian Knigge shows that the tentative counterpart of W24 in 47\,Tuc, 
a possible active binary according to Edmonds et al.\ (2003, Sect.4.5), 
has blue FUV-U colours, which suggests that it is a cataclysmic variable.
At $M_V=2.6$ and $L_x=8.7\times10^{30}$ erg/s, it actually lies below
the line given by Eq.\,\ref{lxmax}.}

\section{Some remarks on theory}

Binaries in a globular clusters change due to their internal evolution
and/or due to external encounters.  To describe the current cluster
binary population one must track the events for each primordial binary
and for each binary that is newly formed via tidal capture, throughout
the cluster.  The first estimates of the formation of binaries with a
neutron star necessarily made a number of drastic simplifications. The
sum of all encounters (of a neutron star with a single star, or with a
binary) was replaced with an integral over the cluster volume of the
encounter rate per unit volume.  Four assumptions followed: the number
density $n_1,n_2$ of each participant in the encounter scales with the
total mass density $\rho$, the relative velocity between the encounter
participants scales with the velocity dispersion $v$, the interaction
cross section $A$ is dominated by gravitational focussing so that
$A\propto1/v^2$, and the encounter rate is dominated by the encounters
in the dense cluster core. Hence one writes the cluster encounter rate
$\Gamma'$ as
\begin{equation}
\Gamma' = \int_V n_1n_2A v dV\propto {\rho_o}^2{r_c}^3/v
\label{gama}\end{equation}
where $\rho_o$ is the central density and $r_c$ the core radius.
If one further eliminates the velocity dispersion through the
virial theorem, $v\propto r_c\sqrt{\rho_o}$, one has
\begin{equation}
\Gamma'  \propto{\rho_o}^{1.5}{r_c}^2\equiv \Gamma
\label{gamb}\end{equation}
where $\Gamma$ is referred to as the collision number.
With a life time $\tau$ the expected number of binaries
of a given type is
\begin{equation}
N=\Gamma'\tau\propto\Gamma\tau
\label{gamc}\end{equation}

A major advantage of these simple estimates is the clear connection
between (the uncertainty in) the input and (the uncertainty of)
the output. Thus, if $n_1,n_2$ are the number densities of neutron stars
and of binaries, respectively, Eqs.\,\ref{gama}-\ref{gamc} indicate
that the uncertainty in the number $N$ of neutron star binaries
scales directly with the uncertainties in $n_1$ and $n_2$.
Similarly, if we overestimate the life time $\tau$ of a binary
by a factor 10, the estimated number $N$ is overestimated
by the same factor.

Thanks to a concerted effort by various groups fairly detailed
computations of the happenings in globular clusters are now
undertaken. This is a fortunate and necessary development, as many
details cannot be understood from the simple scalings above. For
example, the wide progenitors of cataclysmic variables are
destroyed by close encounters before they evolve in a dense cluster
core (Davies 1997), but evolve undisturbed into cataclysmic
variables in the outer cluster regions, from where they can sink to
the dense core to form a significant part of the current
population there (Ivanova et al.\ 2006).

A disadvantage of complex computations is that they tend to hide the
uncertainties. If the cross sections $A$ are described in paper I of a
series, and the life times $\tau$ in paper III, the large
uncertainties in them tend to be less than obvious in paper V where
the final computations are described. The confidence expressed
in summaries of the results of such computations is
sometimes rather larger than warranted. An uncertainty in $A$ or
$\tau$ has an equally large effect in complex computations as in
simple estimates. As a further illustration we discuss
two other uncertainties.

\begin{figure}[]
\parbox[b]{0.54\columnwidth}{\psfig{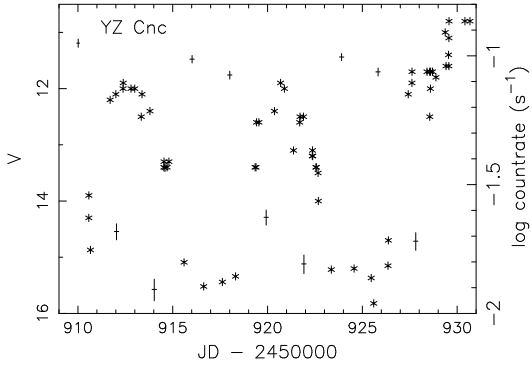}}
\parbox[b]{0.46\columnwidth}{\caption{Optical (*, scale on left) and X-ray 
(+, scale on right) lightcurves of 
the dwarf nova YZ Cnc through several outburst cycles: the outburst in the
optical luminosity is accompanied by a marked {\em drop} in the X-ray 
luminosity. After Verbunt et al.\ (1999)\label{yzcnc}}}
\end{figure}

The first relates to the question what happens when mass transfer from
a giant to its binary companion is dynamically unstable. It is usually
assumed that a spiral-in follows, in which the companion
enters the envelope of the giant and expells it through friction. The
outcome of this process is computed using conservation of energy,
which implies a drastic shrinking of the orbit (Webbink
1984).  However, the study of nearby binaries consisting of two
white dwarfs shows that the mass ratios in them are close to unity
(e.g.\ Maxted et al.\ 2002).  
Such binaries can only be explained if the consequences
of dynamically unstable mass transfer are governed by conservation of
angular momentum, rather than by the energy equation. If the mass
leaving the binary has roughly the same specific angular momentum as
the binary, the orbital period changes relatively little during the
unstable mass transfer and concomitant mass loss from the binary (Van
der Sluys et al.\ 2006). The standard prescription of dynamically
unstable mass transfer hitherto implemented in globular cluster
computations must be replaced.

The second uncertainty relates to the conversion of the mass transfer
rate in a cataclysmic variable to the X-ray luminosity. This
conversion does not affect the evolution of the binary but it is
important for comparison with observations, as most cataclysmic
variables in globular clusters are discovered as X-ray sources. It is
generally assumed that the X-ray luminosity scales directly with the
mass transfer rate: $L_x\propto\dot M$. Alas, reality is more
complicated, and indeed in most cases the X-ray luminosity goes down
when the mass transfer rate goes up. This is demonstrated
unequivocally in dwarf novae whose X-ray luminosity drops
precipitously during outbursts (Fig.\,\ref{yzcnc}), but there is
evidence that it is true in the more stable nova-like variables as
well (Verbunt et al.\ 1997). But exceptions are also known: the dwarf
nova SS Cyg has higher X-ray flux during outburst than in
quiescence (Ponman et al.\ 1995). Theoretical
predictions of the numbers of cataclysmic variables in globular
clusters that radiate detectable X-ray fluxes, are not believable
when based on proportionality of X-ray flux and mass-transfer rate.

Observational evidence for the numbers of binaries of various types,
and of the dependence of these numbers on cluster properties, may
be collected and used to constrain the theories on formation and
evolution of various types of binaries in globular clusters (e.g.\  
Pooley et al.\ 2003, Heinke et al.\ 2006, Pooley \&\ Hut 2006).

In the next Section we describe a new, and we hope more accurate,
method of analysing source numbers: this method is based on
direct application of Poisson statistics. This topic brings one of us, FV,
to a brief Intermezzo.

\section*{Intermezzo by FV: another side of Douglas Heggie}

The first time that I had need of understanding of binomial
and Poisson statistics
was in the early 1980s, when I bought a book by Douglas Heggie (1981)
called {\em Megalithic Science, Ancient Mathemathics and Astronomy
in Northwest Europe}. The first half of this book discusses the
question whether the megalith builders used a standard measure
of length (the answer is no), and the second half studies the
question whether megalithic structures had astronomical orientations.
Whereas Douglas is still rather modest about this book, it is still
{\em the best introduction to the issues in megalithic astronomy} 
(McCluskey 1998, p.11). It, for the first time, explained
the statistics of the study of astronomical orientations of
megalithic structures. In my view it is not an exaggeration
to state that archaeo-astronomy may be divided in a pre-Heggie
era, in which the statistics is usually wrong, and a post-Heggie
era, in which at least some people, following Douglas's prescription,
do their statistics right. Among the latter I may specifically
mention Ruggles (1999).

\begin{figure}
\centerline{
\parbox[b]{0.36\columnwidth}{\psfig{figure=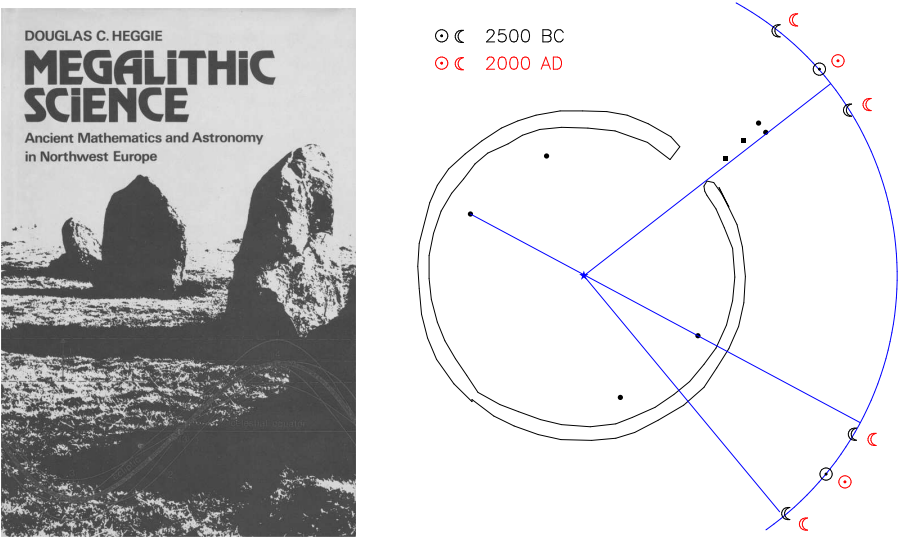,width=0.35\columnwidth,clip=t}}
\parbox[b]{0.64\columnwidth}{\psfig{figure=verbuntf4b.ps,width=0.53\columnwidth,clip=t}}
}
\caption{Left: the classic book by Douglas Heggie on Megalithic Science. 
Right: Sketch of Stonehenge with the circular bank \&\ ditch
(the {\em henge}), the rectangle of the four Station Stones within 
it, and the directions towards the extreme risings of Sun and Moon
at the winter and summer solstitia of 2500 B.C. and 2000 A.D. 
The minute shift in 4500 yr is due to the small change in obliquity.
\label{heggie}}
\end{figure}

As an example of an application of binomial statistics I paraphrase the
discussion on p.198 of the book, of the four Station Stones in
Stonehenge. These Stones were placed, near 2500 B.C., in a rectangle
with the short sides parallel to the main axis of Stonehenge 
(Fig.\,\ref{heggie}).
This main axis points from the center of the rectangle
through a gap in the bank-plus-ditch (the {\em henge}) 
surrounding it toward the location on the horizon of the most
northern annual rising of the Sun, at summer solstice.
The Moon ranges in a 19\,yr cycle from about 5$^\circ$ north of
the Sun to 5$^\circ$ south of the Sun, which leads to a range
of its most northern and southern annual risings. The extremes
of these ranges, together with the two solstitia, define 6 
positions on the eastern horizon. Accepting a range of $\pm1.8^\circ$
for each position, we find a probability that an arbitrary direction
towards the eastern horizon hits one of these positions 
$p=6\times2\times1.8/180=0.12$.
The rectangle of the Station Stones defines three new directions towards
the East: along the long side and along the diagnonals. Two of these
hit one of the 6 positions: extremes of the southernmost annual
setting of the Moon in the 19\,yr cycle.
The probability of 1 chance hit in 3 trials with $p=0.12$ is 28\%, the
probability of 2 chance hits 4\%. Most likely one hit is due to chance,
and one hit intentional.

Thanks to new research at Stonehenge, done after the publication of
{\em Megalithic Science}, we now know that the central structure,
circle and horse-shoe form, were put up simultaneous with the Station
Stones, which implies that the view along the diagonals was blocked.
The intentional hit is therefore the one along the long side. Since
the long side of a rectangle is necessarily perpendicular to the short
side, this implies that the location of Stonehenge was selected for its
latitude.

\section{Testing models against observations}

For the denser clusters it is found that the X-ray sources lie well
within the half-mass radius (e.g.\ NGC\,6440, Fig.1 of 
Pooley et al.\ 2002b), whereas for (apparently) large clusters
only the area within the half-mass radius is covered
(e.g.\ 47 Tuc, Grindlay et al.\ 2001).
Thus, the analysis generally deals with the sources within the
half-mass radius. If $N_h$ is the observed number of X-ray
sources, and $N_b$ the number of background (or foreground)
sources not related to the cluster, the number of cluster
sources within the half-mass radius is $N_c=N_h-N_b$.
From a model we may obtain an estimate of the expected number
of cluster sources $\mu_c$, and we also may estimate the expected
number of background sources $\mu_b$.

In clusters with a small number of sources, $N_c\ltap10$, say, one
cannot apply chi-squared statistics. One way of solving this is by adding
such clusters together, to obtain sufficiently large numbers. This
is done by Pooley \&\ Hut (2006). In this process, the information
on the indicidual clusters is lost.

\begin{figure}
\parbox[b]{0.33\columnwidth}{\psfig{figure=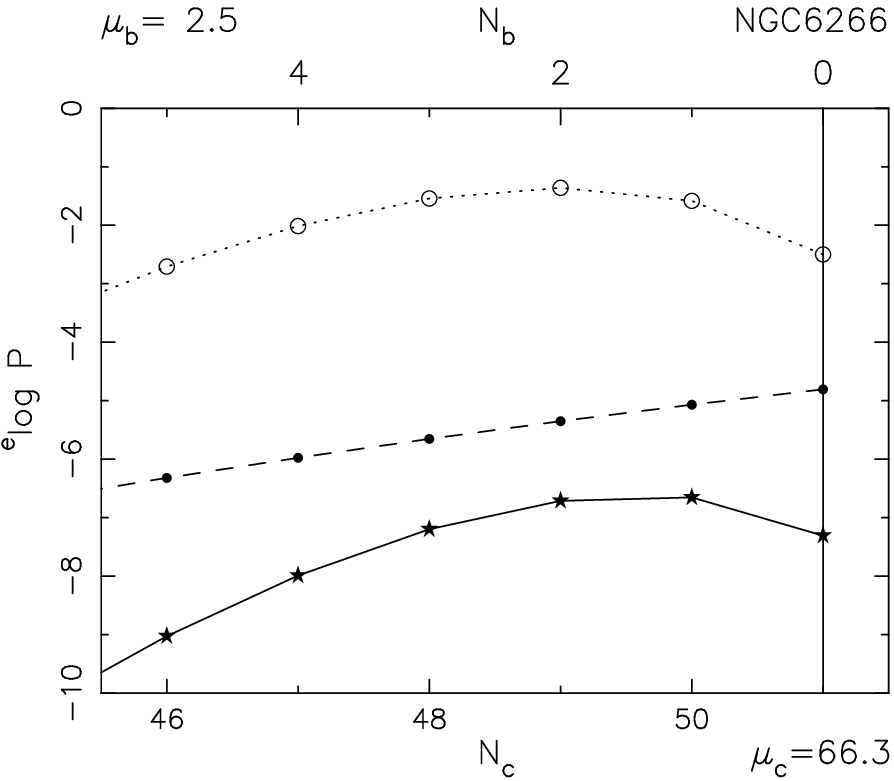,width=0.32\columnwidth,clip=t}}
\parbox[b]{0.33\columnwidth}{\psfig{figure=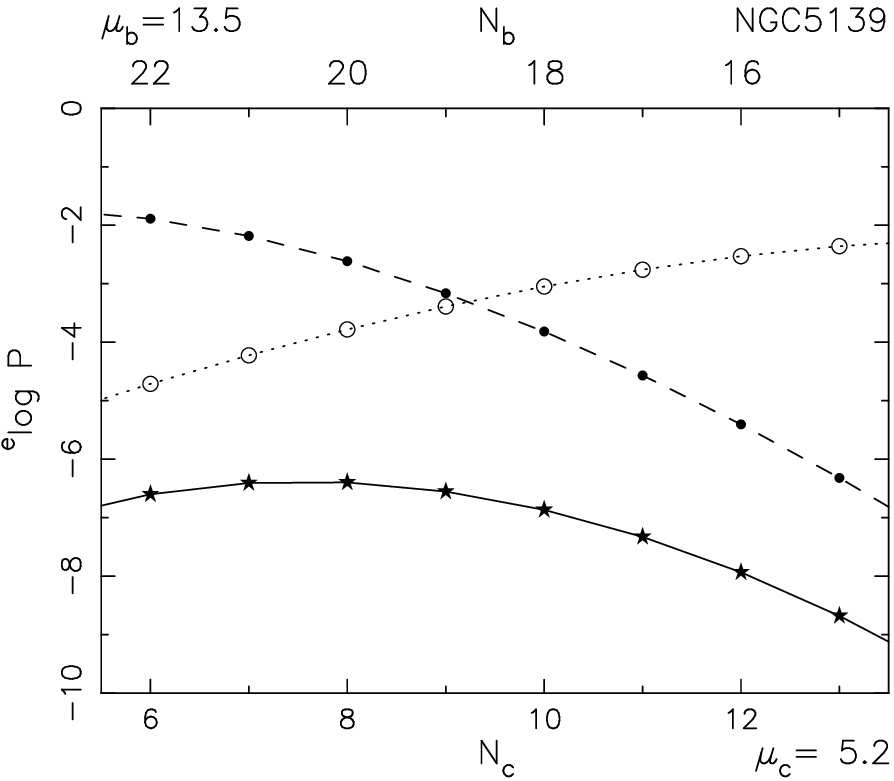,width=0.32\columnwidth,clip=t}}
\parbox[b]{0.33\columnwidth}{\psfig{figure=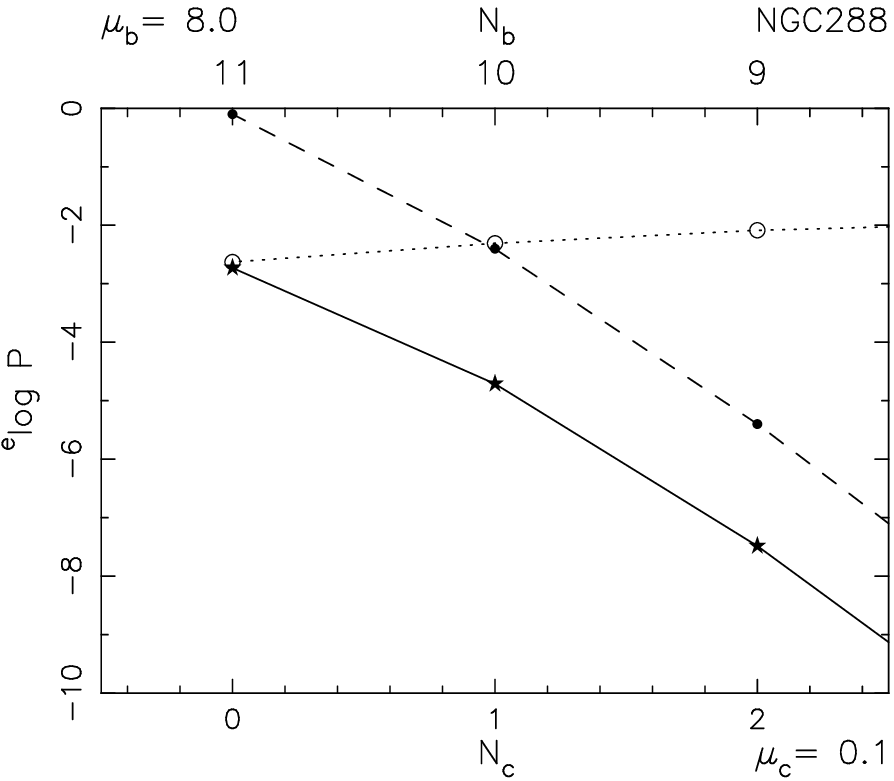,width=0.32\columnwidth,clip=t}}

\caption{Three examples of the combined fit of the number of cluster
sources $N_c$ and background sources $N_b$ for one cluster. 
The expected number of cluster sources $\mu_c$ according to the model
$N_c=1.5\Gamma$, and the expected
number of background sources $\mu_b$ are indicated with each frame.
For each realization $N_c\leq N_h$ the accompanying
realization is $N_b\equiv N_h-N_c$. E.g.\ in NGC\,6266 the observed
number $N_h=51$, allowing the combined realizations of
$N_c,N_b$ as 51,0 or 50,1 or 49,2 etc.
The probability $P(N_c,\mu_c)$ is indicated with a dashed line; 
the probability $P(N_b,\mu_b)$ with a dotted line; and the combined
probability -- the product of these for each allowed pair --
with a solid line. The cases shown are from left to right
for low, roughly equal, and dominant background ($\mu_b\ll N_h$,
$\mu_b\sim 0.5N_h$, $\mu_b\gg\mu_c$).\label{fitex}}
\end{figure}

To avoid this problem, we fit as follows.  The probability of
observing $N$ when $\mu$ is expected according to a Poisson
distribution is
\begin{equation}
P(N,\mu) = {\mu^N\over N!}e^{-\mu}
\end{equation}
An important aspect of the Poison function for our application is
its asymmetry: $P(3,0.1)\ll P(0,3)$, i.e.\ the probability
of observing 3 sources when 0.1 is expected is very much smaller than
the probability of observing 0 sources when 3 are expected.
We now consider
\begin{equation}
N_h = N_c + N_b = \mu_c + \mu_b \qquad \mathrm{with} \qquad 
\mu_c\equiv a\Gamma + b M
\label{modelfit}\end{equation}
Both $N_c$,$N_b$ are realizations of Poisson distributions given by
$\mu_c$ and $\mu_b$ respectively. $\mu_b$ is determined for each
cluster separately, e.g.\ from the observed number of sources well
outside the half-mass radius.  For each value of $N_c$ we take $N_b=N_h-N_c$,
and then compute the combined probability $P(N_c,\mu_c)P(N_b,\mu_b)$.
We then select the $N_c$,$N_b$ pair with the highest combined probability
(Fig.\,\ref{fitex}). The
fitting procedure consists of varying $a$ and $b$ to maximize
\begin{equation}
 P =  \prod_j [P(N_c,\mu_c)P(N_b,\mu_b)]_j
\label{totprob}\end{equation}
where $j$ indexes the clusters. 

\begin{table}
     \begin{minipage}[t]{0.45\columnwidth}
\begin{tabular}{lrrrrr}
cluster  & $\Gamma$\phantom{2} & $M$ & $N_h$ & $\mu_b$ & $N_s$\\
NGC 6266 & 44.2 &  6.3 & 51 &  2.5 \\
NGC 104  & 29.7 &  7.7 & 45 &  4.0 \\
NGC 6626 & 12.6 &  2.5 & 26 &  2.5 \\
NGC 6752 &  6.0 &  1.6 & 11 &  2.5 \\
NGC 7099 &  4.7 &  1.2 &  7 &  1.5 \\
NGC 5904 &  4.1 &  4.4 & 16 &  5.5 \\
NGC 5139 &  3.5 & 17.2 & 28 & 13.5 \\
NGC 6397 &  2.4 &  0.6 & 12 &  0.5 \\
NGC 6121 & $\equiv$1.0 & $\equiv$1.0 &  5 &  2.0 \\
NGC 6809 &  0.2 &  1.4 & 15 &  7.0 & 3\\
NGC 6366 &  0.1 &  0.3 &  5 &  4.0 & 1 \\
NGC 288  &  0.0 &  0.7 & 11 &  8.0 & 2
\end{tabular}
     \end{minipage}
     \begin{minipage}[t]{0.55\columnwidth}\vspace*{-2.4cm}
\caption{Collision number $\Gamma$ and mass $M$, normalized on the
values for NGC\,6121, for clusters studied with Chandra, together with
the number $N_h$ of sources within the half-mass radius, the
expected contribution of background sources $\mu_b$, and the number of
secure members $N_s$. Most values $N_h$ and $\mu_b$ are from Pooley et
al.\ (2003); those for NGC\,288, NGC\,6366 and NGC\,6809 are from
estimated conversion of $L_\mathrm{0.5-2.5\,keV}$ as given in Kong et
al.\ (2006) and Bassa et al.\ (2007) to
$L_\mathrm{0.5-6.0\,keV}$. Collision numbers and masses are derived
from the central density (in $L_\odot$\,pc$^{-3}$), core radii, and
absolute magnitudes given in the Feb 2003 version of the Harris (1996)
compilation, and they are scaled on the values for M\,4 =
NGC\,6121. Thus we assume that the mass to light ratio is the same for
all clusters.
\label{numbers}}
     \end{minipage}
\end{table}

\begin{figure}
\parbox[b]{0.5\columnwidth}{\psfig{figure=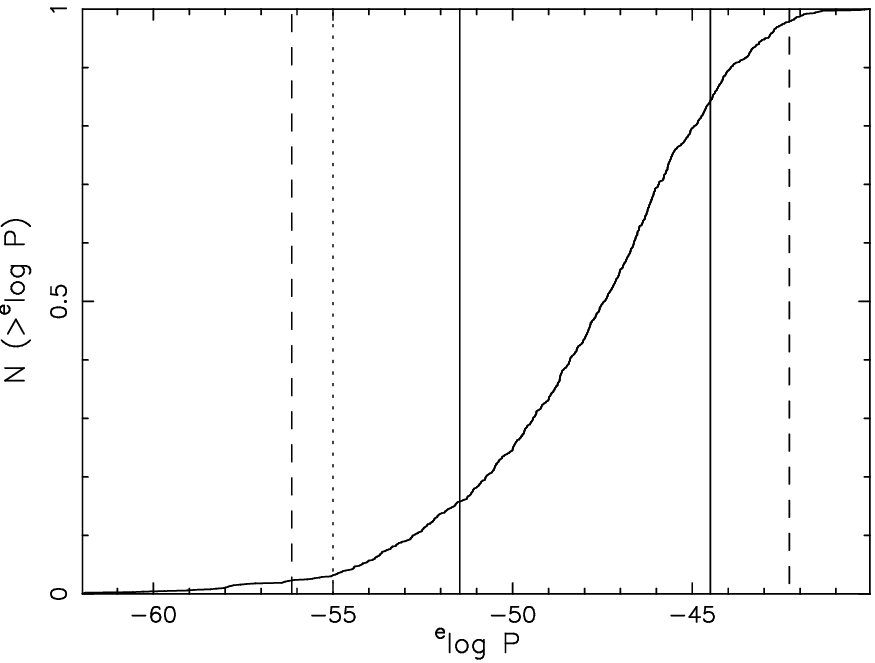,width=0.49\columnwidth,clip=t}}
\parbox[b]{0.5\columnwidth}{\psfig{figure=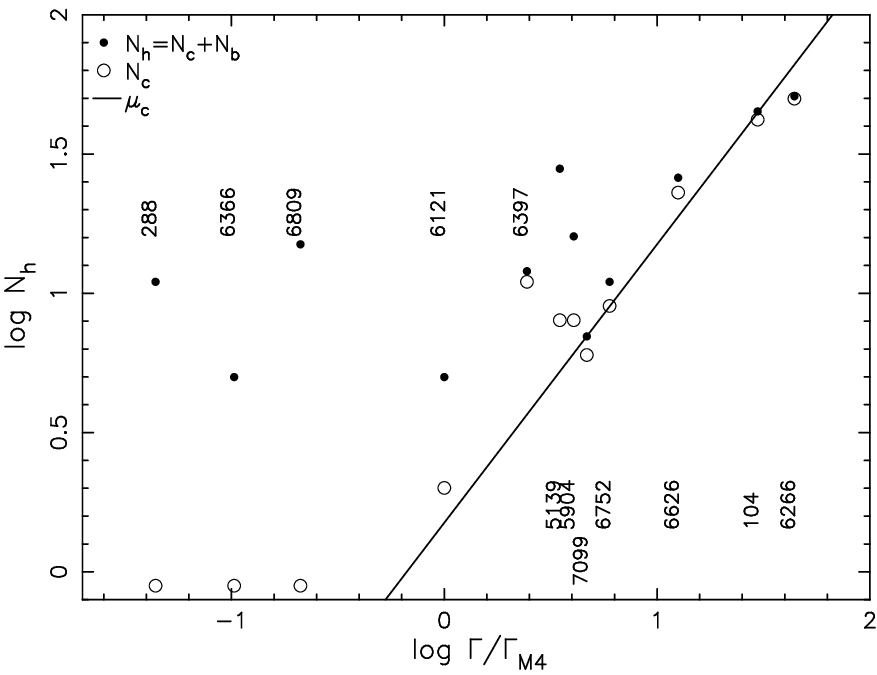,width=0.49\columnwidth,clip=t}}

\caption{Left: distribution of $^e\log P$ for 1000 random realizations
of the best model $\mu_c=1.5\Gamma$, with the 1- and 2-$\sigma$
ranges indicates by solid and dashed lines. The $^e\log P$ value for
the best fit is indicated with a dotted line. Right: The best model
for $\mu_c=1.5\Gamma$ is indicated with a solid line,
the best number $N_c$ with $\circ$ ($<0$ indicates $N_c=0$), 
and the observed number $N_h=N_c+N_b$
with $\bullet$. \label{besta}}
\end{figure}

Table\,\ref{numbers} lists the numbers that we use in our fitting.  We
note from the Table that mass cannot be used as a proxy for collision
number. The cluster with the highest collision number has a ratio
$\Gamma/M\simeq7$ (normalized on M\,4), the cluster with the highest
mass has $\Gamma/M\simeq0.2$. This suggests that mass and collision
number can be well discriminated, and we confirm this below.  The
derived values for $N_h$ and $\mu_b$ (for sources with
$L_\mathrm{0.5-6.0 keV}>4\times10^{30}$ erg/s) correlate with the
derived values of $\Gamma$ and $M$, since all are based on assumed
values for the cluster distance $d$ and interstellar absorption $A_V$.
For the clusters in the Table the correlations thus introduced are
small compared to the ranges of $\Gamma$ and $M$, because the cluster
distances are relatively well known, and we ignore them in what
follows.

The models we fit all have the form given by Eq.\,\ref{modelfit}.
We first fit models with $b=0$, i.e.\ assuming that the number of cluster
sources depends only on the collision number. The best solution has $a=1.5$
and is shown in Figure\,\ref{besta}. The total probability as defined
in Eq.\,\ref{totprob} is $P = e^{-55}$. To see whether this is
acceptable, we have used a random generator to produce 1000 realizations
by drawing random numbers from Poisson distributions with values $\mu_c$ 
as given by the best model and $\mu_b$ as listed in Table\,\ref{numbers}, 
and for
each realization computed the total probability $P$. The cumulative
distribution of $^e\log P$ is also shown in Figure\,\ref{besta}: the
best solution is within the 95\% range around the median value, and thus
acceptable.

\begin{figure}[]
\parbox[b]{0.5\columnwidth}{\psfig{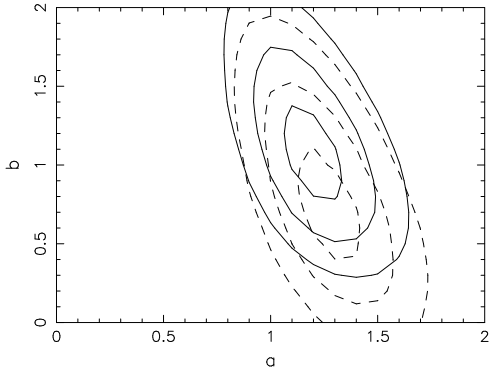}}
\parbox[b]{0.5\columnwidth}{\caption{1,2,3-$\sigma$ contours for the best values
for the model of Eq.\,\ref{modelfit}. Dashed contours are for
the fits which allow all observed sources to be background
sources, solid contours for the fits where a minimum of
secure cluster sources is imposed. \label{fitb}}}
\end{figure}

Our next fit allows non-zero values for $a$ and $b$, i.e.\ allows a
linear dependence both on mass and on collision number. The best solution
now has $a=1.3$ and $b=0.7$; contours of 1,2,3 $\sigma$ around these
best values are shown in Figure\,\ref{fitb}, where we assume that
the distribution of $-2(^e\log P-^e\log P_\mathrm{max})$ is given
by a $\chi^2$ statistic with one degree of freedom. ($P_\mathrm{max}$
is the total probability of the best solution.) We see that the solution
with $b=0$ is marginally acceptable, i.e.\ the evidence for the mass
dependence is marginal. Because we find an acceptable solution
for $\Gamma$ as given by Eq.\,\ref{gamb}, there is no need for
a different dependence on central density.

Our final fit uses more information, {\em viz.} the number of
(almost) certain cluster members among the X-ray sources, as 
determined though optical identifications. The importance of
this may be seen from Fig.\,\ref{fitex} for NGC\,288. With an
estimated background $\mu_b=8$ and, for a model depending on
collision number only, $\mu_c\simeq0.1$, the most probable
solution is that all 11 sources within the half-mass radius
are background sources (cf.\ Fig.\,\ref{besta}). 
If we add the constraint that at least
two sources included in $N_h$ are cluster members (Kong et al.\ 2006)
this solution is no longer possible, and the post probable solution
now has $N_c=2$ and $N_b=9$. Because the collision number is
so small, any cluster source must be a primordial binary,
as indeed argued by Kong et al. The best solution for all clusters
combined now has $a=1.2$ and $b=1.1$, and the solution with $b=0$ now
lies well outside the 3-$\sigma$ contour, i.e.\ the dependence on
mass $M$ is significant (Fig.\,\ref{fitb}).

\section{Conclusions}

\begin{itemize}
\item mass $M$ is {\em not} a proxy for collision number $\Gamma$
\item the number of dim sources scales both with collision number $\Gamma$
and with mass $M$
\item scaling with mass only is not acceptable
\item correct treatment of the background is important, esp.\ for faint
sources
\item to prove the mass-dependence optical identifications are essential
\end{itemize}


\begin{thebibliography}{26}
\expandafter\ifx\csname natexlab\endcsname\relax\def\natexlab#1{#1}\fi

\bibitem[{Bassa {et~al.}(2004)Bassa, Pooley, Homer, Verbunt, Gaensler, Lewin,
  Anderson, Margon, Filippenko, Kaspi, \& van~der Klis}]{bph+04}
Bassa, C., Pooley, D., Homer, L., \& et~al. 2004, ApJ, 609, 755

\bibitem[{Bassa {et~al.}(2007)Bassa, Pooley, Verbunt, \& et~al.}]{bpv+07}
Bassa, C., Pooley, D., Verbunt, F., \& et~al. 2007, A\&A, to be submitted

\bibitem[{Becker {et~al.}(2003)Becker, Swartz, Pavlov, \& et~al.}]{bsp+03}
Becker, W., Swartz, D., Pavlov, G., \& et~al. 2003, ApJ, 594, 798

\bibitem[{Davies(1997)}]{dav97}
Davies, M. 1997, MNRAS, 288, 117

\bibitem[{Dempsey {et~al.}(1993)Dempsey, Linsky, Fleming, \& Schmitt}]{dlfs93}
Dempsey, R., Linsky, J., Fleming, T., \& Schmitt, J. 1993, ApJS, 86, 599

\bibitem[{Edmonds {et~al.}(2003)Edmonds, Gilliland, Heinke, \&
  Grindlay}]{eghg03}
Edmonds, P., Gilliland, R., Heinke, C., \& Grindlay, J. 2003, ApJ, 596, 1177

\bibitem[{Ferraro {et~al.}(2001)Ferraro, Possenti, D'Amico, \& Sabbi}]{fpas01}
Ferraro, F., Possenti, A., D'Amico, N., \& Sabbi, E. 2001, ApJ, 561, L93

\bibitem[{Grindlay {et~al.}(2001)Grindlay, Heinke, Edmonds, \& Murray}]{ghem01}
Grindlay, J., Heinke, C., Edmonds, P., \& Murray, S. 2001, Science, 292, 2290

\bibitem[{Harris(1996)}]{har96}
Harris, W. 1996, AJ, 112, 1487

\bibitem[{Heggie(1981)}]{heggie81}
Heggie, D. 1981, Megalithic Science, Ancient Mathemathics and Astronomy in
  Northwest Europe (London: Thames and Hudson)

\bibitem[{Heinke {et~al.}(2003)Heinke, Edmonds, Grindlay, LLoyd, Cohn, \&
  Lugger}]{heg+03}
Heinke, C., Edmonds, P., Grindlay, J., \& et~al. 2003,
  ApJ, 590, 809

\bibitem[{Heinke {et~al.}(2005)Heinke, Grindlay, Edmonds, \& et~al.}]{hge+05}
Heinke, C., Grindlay, J., Edmonds, P., \& et~al. 2005, ApJ, 625, 796

\bibitem[{H{\"u}nsch {et~al.}(1999)H{\"u}nsch, Schmitt, Sterzik, \&
  Voges}]{hssv99}
H{\"u}nsch, M., Schmitt, J., Sterzik, M., \& Voges, W. 1999, A\&AS, 135, 319

\bibitem[{Ivanova {et~al.}(2006)Ivanova, Heinke, Rasio, \& et~al.}]{ihr+06}
Ivanova, N., Heinke, C., Rasio, F., \& et~al. 2006, MNRAS, 372, 1043

\bibitem[{Kong {et~al.}(2006)Kong, Bassa, Pooley, \& et~al.}]{kbp+06}
Kong, A., Bassa, C., Pooley, D., \& et~al. 2006, ApJ, 647, 1065

\bibitem[{Maxted {et~al.}(2002)Maxted, Marsh, \& Moran}]{mmm02}
Maxted, P., Marsh, T., \& Moran, C. 2002, MNRAS, 332, 745

\bibitem[{McCluskey(1998)}]{mccluskey98}
McCluskey, S. 1998, Astronomies and cultures in early medieval {E}urope
  (C.U.P.)

\bibitem[{Ponman {et~al.}(1995)Ponman, Belloni, Duck, Verbunt, Watson,
  Wheatley, \& Pfeffermann}]{pbd+95}
Ponman, T., Belloni, T., Duck, S., \& et~al. 1995, MNRAS, 276, 495

\bibitem[{Pooley \& Hut(2006)}]{ph06}
Pooley, D. \& Hut, P. 2006, ApJ, 646, L143

\bibitem[{Pooley {et~al.}(2003)Pooley, Lewin, Anderson, \& et~al.}]{pla+03}
Pooley, D., Lewin, W., Anderson, S., \& et~al. 2003, ApJ, 591, L131

\bibitem[{Pooley {et~al.}(2002{\natexlab{a}})Pooley, Lewin, Homer, Verbunt,
  Anderson, Gaensler, Margon, Miller, Fox, Kaspi, \& van~der Klis}]{plh+02}
Pooley, D., Lewin, W., Homer, L., \& et~al. 2002{\natexlab{a}}, ApJ, 569, 405

\bibitem[{Pooley {et~al.}(2002{\natexlab{b}})Pooley, Lewin, Verbunt, Homer,
  Margon, Gaensler, Kaspi, Miller, Fox, \& van~der Klis}]{plv+02}
Pooley, D., Lewin, W., Verbunt, F., \& et~al. 2002{\natexlab{b}}, ApJ, 573,
  184

\bibitem[{Ruggles(1999)}]{ruggles99}
Ruggles, C. 1999, Astronomy in prehistoric {Britain} and {I}reland (New Haven:
  Yale Univ. Press)

\bibitem[{van~der Sluys {et~al.}(2006)van~der Sluys, Verbunt, \& Pols}]{svp06}
van~der Sluys, M., Verbunt, F., \& Pols, O. 2006, A\&A, 460, 209

\bibitem[{Verbunt(2001)}]{ver01}
Verbunt, F. 2001, A\&A, 368, 137

\bibitem[{Verbunt {et~al.}(1997)Verbunt, Bunk, Ritter, \& Pfeffermann}]{vbrp97}
Verbunt, F., Bunk, W., Ritter, H., \& Pfeffermann, E. 1997, A\&A, 327, 602

\bibitem[{Verbunt \& Hut(1987)}]{vh87}
Verbunt, F. \& Hut, P. 1987, in The Origin and Evolution of Neutron Stars, IAU
  Symposium No. 125, ed. D.~Helfand \& J.-H. Huang (Dordrecht: Reidel),
  187--197

\bibitem[{Verbunt \& Lewin(2006)}]{vl06}
Verbunt, F. \& Lewin, W. 2006, in Compact stellar {X}-ray sources, ed. W.~Lewin
  \& M.~van~der Klis (Cambridge University Press), 341--379

\bibitem[{Verbunt {et~al.}(1999)Verbunt, Wheatley, \& Mattei}]{vmw99}
Verbunt, F., Wheatley, P., \& Mattei, J. 1999, A\&A, 346, 146

\bibitem[{Webb {et~al.}(2006)Webb, Wheatley, \& Barret}]{wwb06}
Webb, N., Wheatley, P., \& Barret, D. 2006, A\&A, 445, 155

\bibitem[{Webbink(1984)}]{web84}
Webbink, R. 1984, ApJ, 277, 355

\bibitem[{Wijnands {et~al.}(2002)Wijnands, Heinke, \& Grindlay}]{whg02}
Wijnands, R., Heinke, C., \& Grindlay, J. 2002, ApJ, 572, 1002

\end{thebibliography}
\end{document}